\documentclass[twocolumn,preprintnumbers,amsmath,amssymb]{revtex4}

\usepackage{graphicx}
\usepackage{dcolumn}
\usepackage{bm}
\usepackage{subfigure}

\begin{document}

\title{Crescent Waves in Optical Cavities}

\author{Chandroth P. Jisha, YuanYao Lin, Tsin-Dong Lee and Ray-Kuang Lee}
 \affiliation{
Institute of Photonics Technologies, National Tsing-Hua University, Hsinchu, 300, Taiwan
}

\date{\today}
\begin{abstract}
We theoretically and experimentally generate stationary crescent surface solitons  pinged to the boundary of a micro-structured vertical cavity surface emission laser by using the intrinsic cavity mode as a background potential.
Instead of a direct transition from linear to nonlinear cavity modes, we demonstrate the existence of a symmetry-breaking crescent waves without any analogs in the linear limit.
Our results provide  an alternative and general method to control lasing characteristics as well as to study optical surface waves.
\end{abstract}


\maketitle
By state-of-the-art semiconductor technologies, microcavities have provided a controllable confinement and manipulation of photons with small mode volumes and  ultrahigh quality factors~\cite{RKChang, Vahala} .
For an integrable cavity shape,  supported  stable resonance modes  have attracted much attention in photonics, quantum electrodynamics, and telecommunications, due to their potential applications to modulate spontaneous emission and make thresholdless lasing~\cite{Vahala-nature-review}.
If the cavity shape is slightly deformed into a chaotic one, unstable localized waves, coined as scar modes, are found also to support lasing, which produce unidirectional outputs and provide an alternative understanding in the correspondence between classical and quantum system~\cite{Nockel, Gmachl, scarTD2008, deformTD2009}.

To form standing light waves, optical cavities are typically defined by the geometry of media with a higher refractive index.
Confined by total internal reflection at the interface,  whispering-gallery modes are almost grazing incidence patterns, which act as filters, delay lines, couplers, and sensors for a broad areas from optical communications, information processing, to biophotonics~\cite{WGM, APB09}.
Instead of the cavity modes with periodic orbits, light wave can attach to the boundary of materials under appropriate conditions~\cite{Pochi, ZChen, plasmon}.
As electrons localized at crystalline surfaces known as Tamm and Shockley states ~\cite{Tamm, Shockley}, 
optical surface waves are localized at the interface between two different media.
Direct observations of optical surface states have been demonstrated in photonic lattice edges~\cite{discrete_surface} and periodic waveguide arrays~\cite{waveguidarray}.

Without introducing any symmetry-breaking in the geometry, optical surface waves can also be induced through optical nonlinearity, resulting in the formation of surface solitons~\cite{Torner-09}. 
These surface solitons  are supported even in uniform materials, and can not find any analogs in the linear cases.  
A specific type of localized states, in the form of {\it crescent surface solitons},  is pinged to a circular boundary with a similar shape as Barchan sand dunes~\cite{veit:nature:2003}. 
Optical crescent solitons have been predicted theoretically to exist in highly nonlocal media \cite{highly}, through the superposition of two vortex beams with  different topological charges~\cite{topology}, or by introducing inhomogeneous losses~\cite{loss}.
In rotating Bose-Einstein condensates, matter-wave crescent vortex solitons are found in the 2D Gross-Pitaevskii equation combining the local self-attractive nonlinearity and a quadratic-quartic potential~\cite{BEC}. 
With a structure of concentric rings in the refractive index modulation, stationary and rotating surface solitons  with thresholdless formation powers are shown to exist at the edge of guiding structures~\cite{Tornor-rotate}.

In this Letter we demonstrate a new type of microcavities, fabricated on the surface of a Vertical Cavity Surface Emission Laser (VCSEL), and report experimentally the formation of crescent surface solitons by collecting near-field radiation intensities.
By increasing the injection current, we analyze transitions between linear Laguerre-Gaussian-like cavity modes and nonlinear optical patterns resembling soliton rings .
In contrast to a direct crossover between linear and nonlinear modes,  we introduce the concept to design the intrinsic linear cavity modes as a background potential, which near the threshold lasing condition supports stationary single-, double-, and quadruple-humped crescent solitons without any counterparts in the linear limit.
Our numerical results  based on a  nonlinear wave equation within a focusing medium
are in a good agreement with the experimental  observations.
The experimental and numerical investigations in this work provide an effective state for investigating surface waves in micro-structured semiconductor lasers.

\begin{figure}[h]
\centering
\includegraphics[width=8.3cm]{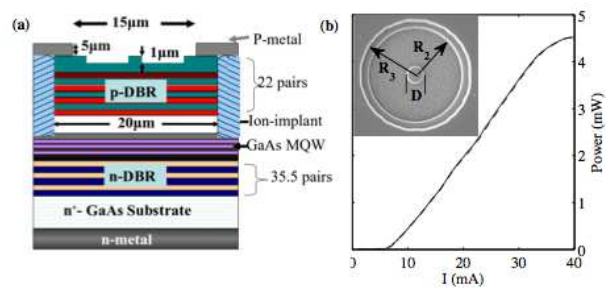}
\caption{\label{fig:fig1}(Color online) (a) Schematic diagram of the device structure. (b) The L-I curve, power versus current, for our VCSEL with a surface microstructure.
The inset shows the top-view SEM image, with the geometric
diameter $D$ = $2\mu$m, two radius $R_2$ = $6.5\mu$m and $R_3$ = $7.5\mu$m.}
\end{figure}

\begin{figure}[ht]
\centering
\includegraphics[width=8.3cm]{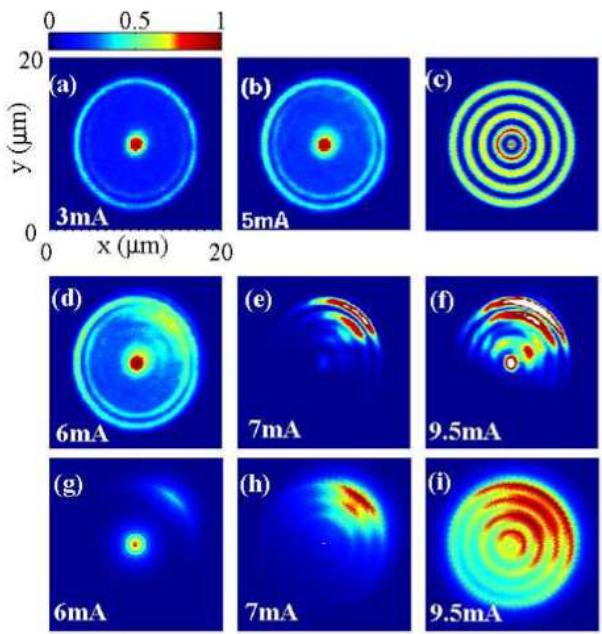}
\caption{\label{fig:fig2}(Color online) Experimental demonstration of a mode transition for crescent surface waves with the injection currents for  (a) $3$mA, (b) $5$mA, (d) $6$mA, (e) $7$mA, and (f) $9.5$mA, respectively. (c) shows the corresponding numerical result for the supported linear $LG_4^0$-like cavity mode. Nonlinear crescent surface modes
with (g) single-, (h) double-, and (i) quadruple-humps are shown with the same simulation parameters but just varying the injection current.}
\end{figure}
\begin{figure}
\centering
\includegraphics[width=8.3cm]{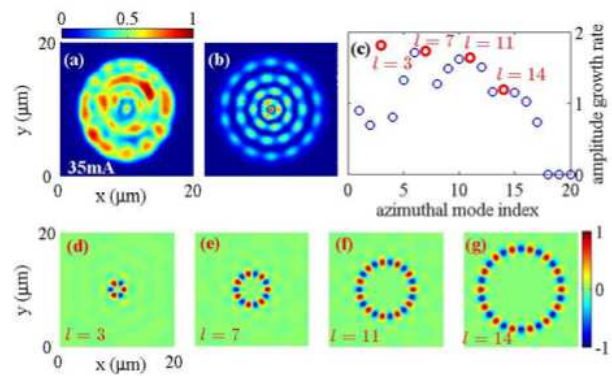}
\caption{\label{fig:fig3}(Color online) (a) Experimental demonstration of the mode pattern resembling a soliton-ring structure at  the injection current $35$mA. (b) shows the numerical result for a $4^{th}$-order nonlinear cylindrically symmetric mode composited by the superposition of unstable eigen-modes with azimuthal indexes of (d) $l=3$, (e) $l=7$, (f)$l=11$, and (g) $l=14$, respectively. (c) shows the corresponding azimuthal instability spectrum.}
\end{figure}

Our optical cavity is fabricated on an ion-implanted VCSEL, with the schematic diagram  shown in Fig. 1(a).
The epitaxial layers of the VCSELs are grown by metal organic chemical vapor deposition (MOCVD) on a $n^+$-GaAs substrate, with a graded-index separate confinement heterostructure (GRlNSCH) active region formed by undoped triple-GaAs-AlGaAs quantum wells placed in one lambda cavity, where the emitting window is designed as $15\mu$m in diameter and the emitting aperture is  confined by the implanted region as  $20\mu$m in diameter.
On the surface, we etch the emitting window  by the focus ion beam (FIB) to $1\mu$m in depth with a donut-shape mask of  $1\mu$m inner radius and  $6.5\mu$m outer radius.
From the  scanning electron microscope (SEM) image, the inset in Fig. 1(b),   top view of the surface structure displays that in the center there is a circular region with a higher refractive index,  surrounded by the etched donut-shape medium with a lower refractive index,  and then comes the outer annular cavity with a higher refractive index modulation again. 
The L-I curve, light versus current, of the surface-structured VCSEL is shown in Fig. 1(b).
The threshold current of the device is about $6$mA with the lasing wavelength at $855$nm.
It would be demonstrated later, this kind of cavity geometry can generate crescent surface waves pinging to the outer circular boundary effectively. 

Next, we measure the near-field electromagnetic intensity distribution at a fixed injection current by a charge-coupled device (CCD) camera through a standard microscope with a $100X$ lens. 
While the VCSEL is operated below threshold, for example, at the current of $3$mA, it can be seen clearly in Fig. 2(a) that spontaneous emission pattern just reflects the geometry of our optical cavity in the VCSEL.
Since the circular cavity in the central region has a smaller mode volume and a higher quality factor, as the injection current increases, it is clearly to see that the light intensity in the central region becomes brighter and brighter, as shown in Fig. 2(b).
After the VCSEL is turned on, our device begins to lase.
Instead of a single central spot, a series of  surface wave patterns resembling crescent surface modes are presented in  Figs.~\ref{fig:fig2}(d-f) for the injection currents of $6$, $7$ and $9.5$mA,  respectively. 
Even though our optical cavity is designed with a symmetric circular geometry,  
these localized surface modes have a symmetry-breaking shape along the azimuthal direction.
Moreover, these lasing modes are generated at the outer circular boundary first, then multiple-humped profiles in the shape of  crescent waves appear regularly in the etched donut-shape region.
When the operation current is selected well above the threshold, nonlinearity becomes a dominant effect in the formation of the transverse optical patterns, such as $35$mA shown in Fig.~\ref{fig:fig3}(a).
The observed structure itself is believed to resemble a cluster of soliton rings~\cite{elliptical}.
This is confirmed by the subsequent theoretical analysis summarized below.

The complex spatiotemporal dynamics and the pattern formation in a broad-area semiconductor laser cavity have been theoretically analyzed to study  filamentary behaviors, soliton manipulations, and transverse mode transitions.
To illustrate theoretically the formation of observed crescent surface modes in a semiconductor laser cavity, we start with the coupled equations for the slowly varying complex envelope of the electric field $E$ and carrier density $N$~\cite{model}:
\begin{eqnarray*}
&&\partial_t E= -(1+\eta+i\theta)E-i2C\Theta(N-1)E+i\,{\partial_\perp}^2 E, \label{eq:csE}\\
&&\partial_t N=-\gamma\left[N+\beta N^2 - I + |E|^2(N-1) - d\,{\partial_\perp}^2 N \right], \label{eq:csN}
\end{eqnarray*}
where $C$ is the saturable absorption coefficient scaled to the resonator transmission, ${\partial_\perp}^2$ is the transverse Laplacian describing the diffraction in the paraxial approximation, $\eta$ is the linear absorption coefficient due to the material in the regions between the semiconductor and the mirrors, $\theta$ is the cavity detuning, $\Theta=(i+\alpha)$ represents the absorptive response of material,  $\alpha$ is the corresponding line-width enhancement factor, and  $d$ is the diffusion constant of the carrier scaled to the diffraction coefficient. $\gamma$ and $\beta$ are the normalized decay rates of the carrier density that describe the non-radiative and radiative carrier recombinations, respectively. External injection current is denoted by $I$.

To find spatially localized solutions numerically, we assume that the diffusion length of carriers is much smaller than the diffraction length of the electromagnetic field,
i.e., $d\ll 1$,  also neglecting the second-order effects in the evolution of the carrier density by setting $\beta=0$. When the system reaches equilibrium, as it happens in our laser cavity, the field $|E|^2$ becomes stationary and we can replace $N-1$ to the first order in Fourier space~\cite{yylin2007}, i.e.,
\begin{eqnarray*}
N-1 \approx \int_{-\infty}^t \mathrm{exp}\left[-\gamma(1+|E|^2-d\partial_x^2)(t-t')\right]\gamma(I-1)dt'.
\end{eqnarray*}
By neglecting the variation in the coefficient of differential operator and assuming $d$ constant, we derive a reduced dissipative wave equation for the
electromagnetic waves in the semiconductor microcavity \cite{nlin},
\begin{eqnarray}
&& i\partial_\tau E + \delta\bar{\theta}E - (1+i\delta)\left[ -(1+\eta)+\frac{2C(I-1)}{1+|E|^2} \right]E \nonumber\\
&& + (1-i\delta{d})\partial_\perp^2 E  + V(r,\theta) E =0,
\label{eq:dnse2}
\end{eqnarray}
where $\tau=\alpha t$ and the spatial coordinates is normalized with a factor $\sqrt{\alpha}$. We also define $\bar{\theta}=\theta-\theta_0$ for $\theta_0+\alpha(1+\eta)=0$, and $\delta=1/\alpha$ to simplify the notations.
The last term in Eq. (\ref{eq:dnse2}), $V(r, \theta)$, is added to account the refractive index modulation for  the cavity geometry.
When the line-width enhancement factor $\alpha\gg 1$ ($\delta\ll 1$), the proposed equation (\ref{eq:dnse2}) can be approximated by the generalized nonlinear Schr{\"o}dinger equation, where modulation instability is known to lead to the formation of nonlinear patterns and solitons~\cite{ccjeng2009}.

Before applying the nonlinear wave model, we define the index modulation based on the  lateral geometry measured experimental in Fig.~\ref{fig:fig2}(a) by assuming the
effective refractive indexes as $3.49$ and $2$ for the regions  without and with the FIB etch process, respectively.
The calculated linear eigenmodes (TE modes) of this cavity geometry with the lasing wavelength are found to support two kinds of cavity modes; one has a $4$-rings profile very similar to a Laguerre-Gaussian(LG) mode of $4^{th}$-order in the radial and $0^{th}$ in the azimuthal directions, denoted by $LG_{4}^{l=0}$, for which the field distribution occupies most area of the whole cavity, as shown in Fig.~\ref{fig:fig2}(c); while the other one has a single peak just inside the central region that is close to a $LG_{0}^{0} $.
These modes are typical eigenmodes in the optical fiber waveguides and circular resonators.

The central region in this micro-structured geometry acts as an auxiliary cavity, of which the size plays a crucial role in our design to generate crescent surface modes.
In the limit of a geometry without this central cavity, our VCSEL is a broad-area circular cavity, with an annual index modulation in the outer ring.
Although this type of a broad-area cavity also supports radially distributed linear modes as $LG_{4}^{0}$-like mode, but practically it is hard to produce a symmetry-breaking mode as crescent waves due to the mode competitions.
On the other hand, if the size of the central region is larger, we found experimentally, not shown here, that only lasing in the central cavity is possible and the lasing in the outer region is totally suppressed.

Due to the fact that the intrinsic $LG_{4}^{0}$-like mode in Fig.~\ref{fig:fig2}(c) is suppressed as a result of mode competition with the single-peak mode in Fig.~\ref{fig:fig2}(a), we take this $LG_{4}^{0}$-like mode, denoted by $E_L(r, \theta)$ for the field distribution, as a background potential and solve nonlinear wave equation described in Eq. (\ref{eq:dnse2}) with  an additional self-induced refractive index modulation $V_L(r, \theta) = |E_L(r,\theta)|^2$:
\begin{eqnarray}
\label{gov:eq}
\frac{2C(I-1)}{1+|E|^2} \Delta E + \partial_\perp^2 \Delta E  + [V(r,\theta)+V_L(r, \theta)] \Delta E =0.
\end{eqnarray}
Here we treat the electrical field $E(r, \theta)  = E_L + \Delta E$ and find the solution for $\Delta E$ self-consistently.
The validity of this perturbation approach can only be applied to the case near the threshold condition for the reason that the unperturbed electrical field $E_L(r, \theta)$ is suppressed below the lasing condition, but with a perturbed electrical field $\Delta E(r, \theta)$ one has a chance to overcome the threshold and produce a lasing mode.

The simulation results based on Eq. (\ref{gov:eq}) are shown in Fig.~\ref{fig:fig2}(g-i), which successfully report  nonlinear crescent surface modes with single-, double-, and quardruple-humped solutions by using the same parameters but just varying the injection current. %
As mentioned before, with the assistant of  the small cavity in the center, which suppress the supported $LG_{4}^{0}$-like mode at the first.
As the injection currents increase, this supported $LG_{4}^{0}$-like mode is driven by a higher gain and has to turn in lasing too. %
But before this $LG_{4}^{0}$-like mode to become a dominant lasing mode over the single-peak one, it acts as an effective index modulation in the shape of concentric rings, which support stationary crescent surface modes as predicted theoretically~\cite{Tornor-rotate}. %
A one-to-one correspondence with experimental data in Fig.~\ref{fig:fig2}(d-f) verifies our numerical investigations. %

Such a perturbed approach breaks down as the injection current increases to a certain value. For well above the threshold condition, we solve the nonlinear mode in Eq. (\ref{eq:dnse2}) directly  with the supported linear mode in Fig.~\ref{fig:fig2}(c) as an initial ansatz. %
The convergent mode solutions, which forms a $4^{th}$-order cylindrically symmetric bound-state, are found numerically by the standard relaxation method. %
Owing to the azimuthal instability of cylindrically symmetric higher-order bound states, these ring structures are prone to break into spots lying on the concentric rings, as illustrated in Fig.~\ref{fig:fig3}(b). %
The number of lobes is quite different to each other on distinct rings because the reason that modes with  different azimuthal indexes may have an equivalent  growth rate of instabilities.
As an example, we calculate the corresponding azimuthal instability spectrum for different azimuthal mode indexes in Fig.~\ref{fig:fig3}(c), which gives $4$ possible nonlinear modes with the largest amplitude growth rates, i.e., $l = 3, 7, 11$, and $14$, respectively.
By the superposition of these four nonlinear patterns illustrated in  Fig.~\ref{fig:fig3}(d-g), we find that not only the number of rings but also the number of lobes in the outer ring are in a good agreement for the case of a highly nonlinear pattern experimentally and numerically, as the comparison shown in  Fig.~\ref{fig:fig3}(a) and (b).
In this case, one can see that the $LG_{4}^{0}$-like mode is dominant in the radiation pattern, which later becomes a nonlinear mode with broken spots in the ring-structure as a manifold of nonlinear instability for soliton rings~\cite{elliptical}.

Before conclusion, we would like to address the comparison of our introduced laser cavity to those with an annular Bragg resonators~\cite{circular-Bragg, annual}, where the cavity modes are confined to propagate azimuthally within a cylindrical Bragg grating.
Even though lasing modes within a multiple-ring structure, such as annual Bragg resonators, can be excited easily, but  the mode competition among supported linear cavity modes results in the preference of a symmetry maintaining radiation pattern.
On the contrary, the approach we used here seems unnatural at a first glance for assuming a refractive index modulation induced by the supported linear cavity modes.
As long as one operates near the threshold condition, our proposed using intrinsic non-lasing cavity modes to explain the generation of crescent surface waves works.
It turns out that the preformed symmetry-breaking crescent waves on top of a supported linear cavity modes give us a satisfied comparison with experiments not only qualitatively but also quantitatively.
The results we showed theoretically and experimentally here can also be useful for  studying  a variety of  microcavities with shapes from micro-disks, micro-rings, and micro-spheres.


In summary, by introducing the concept to design an intrinsic non-lasing cavity mode as a background potential, we fabricated a  surface-structured VCSEL and reported the observation of crescent surface waves  near the threshold lasing condition.
A mode transition from  stationary single-, double-, and quardruple-humped crescent solitons to soliton rings are demonstrated numerically and experimentally.
Without any counterparts in the linear limit, the experimental observations and the simulation results provide an alternative but effective approach to access optical surface modes in semiconductor lasers.

\end{document}